\pdfoutput=1 

\documentclass{pnastwo}
\usepackage{graphics,amssymb}

\newcommand{\fig}[5]
{
\begin{figure}[#4]
\begin{center}
\resizebox{#5 \columnwidth}{!}{\includegraphics{#1}}
\caption{\label{#3}#2}
\end{center}    
\end{figure}
}

\newcommand{\figstar}[4] 
{
\begin{figure*}[#4]
\begin{center}
\resizebox{1.90 \columnwidth}{!}{\includegraphics{#1}}
\caption{\label{#3}#2}          
\end{center}        
\end{figure*}
}

\begin{document}

\title{Molecular random tilings as glasses}

\author{Juan P. Garrahan\affil{1}{School of Physics and Astronomy, University of Nottingham, Nottingham, NG7 2RD, U.K.}, Andrew Stannard\affil{1}{}, Matthew O. Blunt\affil{1}{} and Peter H. Beton\affil{1}{}}

\contributor{Submitted to Proceedings of the National Academy of Sciences
of the United States of America}

\maketitle

\begin{article}

\begin{abstract}
We have recently shown [Blunt, MO, et al. (2008) {\em Random tiling and topological defects in a two-dimensional 
molecular network}. Science, 322:1077Ð-1081] that p-terphenyl-3,5,3',5'-tetracarboxylic acid adsorbed on graphite self-assembles into a two-dimensional rhombus random tiling.  This tiling is close to ideal, displaying long range correlations punctuated by sparse localised tiling defects.  In this paper we explore the analogy between dynamic arrest in this type of random tilings and that of structural glasses.  We show that the structural relaxation of these systems is via the propagation--reaction of tiling defects, giving rise to dynamic heterogeneity.  
We study the scaling properties of the dynamics, and discuss connections with kinetically constrained models of glasses. 
\end{abstract}


\abbreviations{KCM, kinetically constrained model; STM, scanning tunneling microscope; TPTC,
p-terphenyl-3,5,3',5'-tetracarboxylic acid}

\section{Introduction}

Figure 1 shows 
a molecular network formed by organic molecules 
adsorbed from solution onto a graphite substrate \cite{Blunt}.   
The molecule, p-terphenyl-3,5,3',5'-tetracarboxylic acid, or TPTC (see Fig.\ 1A), binds to other TPTC molecules on the substrate adopting one of three possible orientations.  Each molecule can then be mapped to a rhombus tile, see Fig.\ 1B, where the colours red, green and blue indicate the molecular orientation.  Neighbouring molecules (tiles) can bind to neighbours in a parallel or ``arrowhead'' configuration, equivalent to junctions between tiles of the same or different colour, respectively.  Fig.\ 1C shows a scanning tunneling microscope (STM) image of the resulting molecular network of adsorbed TPTC, and Fig.\ 1D the corresponding rhombus tiling \cite{Blunt} where each molecule is represented by a tile.  

The molecular networks studied in \cite{Blunt} are close to ``perfect'' rhombus tilings (or dimer coverings of the honeycomb lattice) \cite{Fisher-Kasteleyn,Blothe,Henley}, in the sense that they contain rather few tiling defects, typically less than one defect per 300 adsorbed molecules \cite{Blunt}.  In Figs.\ 1C and 1D one such tiling defect is identified.  They are also entropically stabilized ``random tilings'' displaying algebraic spatial correlations \cite{Blunt}, characteristic of a critical, or Coulomb, phase \cite{Fisher-Kasteleyn,Blothe,Henley}.  The structures such as those of Fig.\ 1 are close to dynamically arrested at room temperature \cite{Blunt}.  The interaction energy between two neighbouring molecules is several times $k_B T$ \cite{Blunt}, so once a tiling is formed tile removal is highly suppressed, and structural relaxation is slow.  Tile rearrangements mediated by propagation of defects have been observed experimentally, but so far on timescales of seconds for each event \cite{Blunt}. This combination of an amorphous structure, albeit with critical spatial correlations, and very slow relaxation suggests an analogy between this kind of random tilings and structural glass formers \cite{reviews}.  

The aim of this paper is to discuss this analogy.  For simplicity we focus on the case where all tile-tile interactions are equal, since the dynamics for small bias \cite{bias} is qualitatively the same, and on the long-time dynamical regime; the initial growth dynamics is discussed in a separate paper \cite{Stannard}.  We show, by means of numerical simulations, that the low temperature dynamics of a rhombus tiling where the number of tiles is not conserved displays some of the features observed in the dynamics of structural glass formers, in particular dynamic heterogeneity.  Relaxation in these random tilings is facilitated by tiling defects, a mechanism similar to that of kinetically constrained models of glasses.  We will also discuss this connection.

\fig{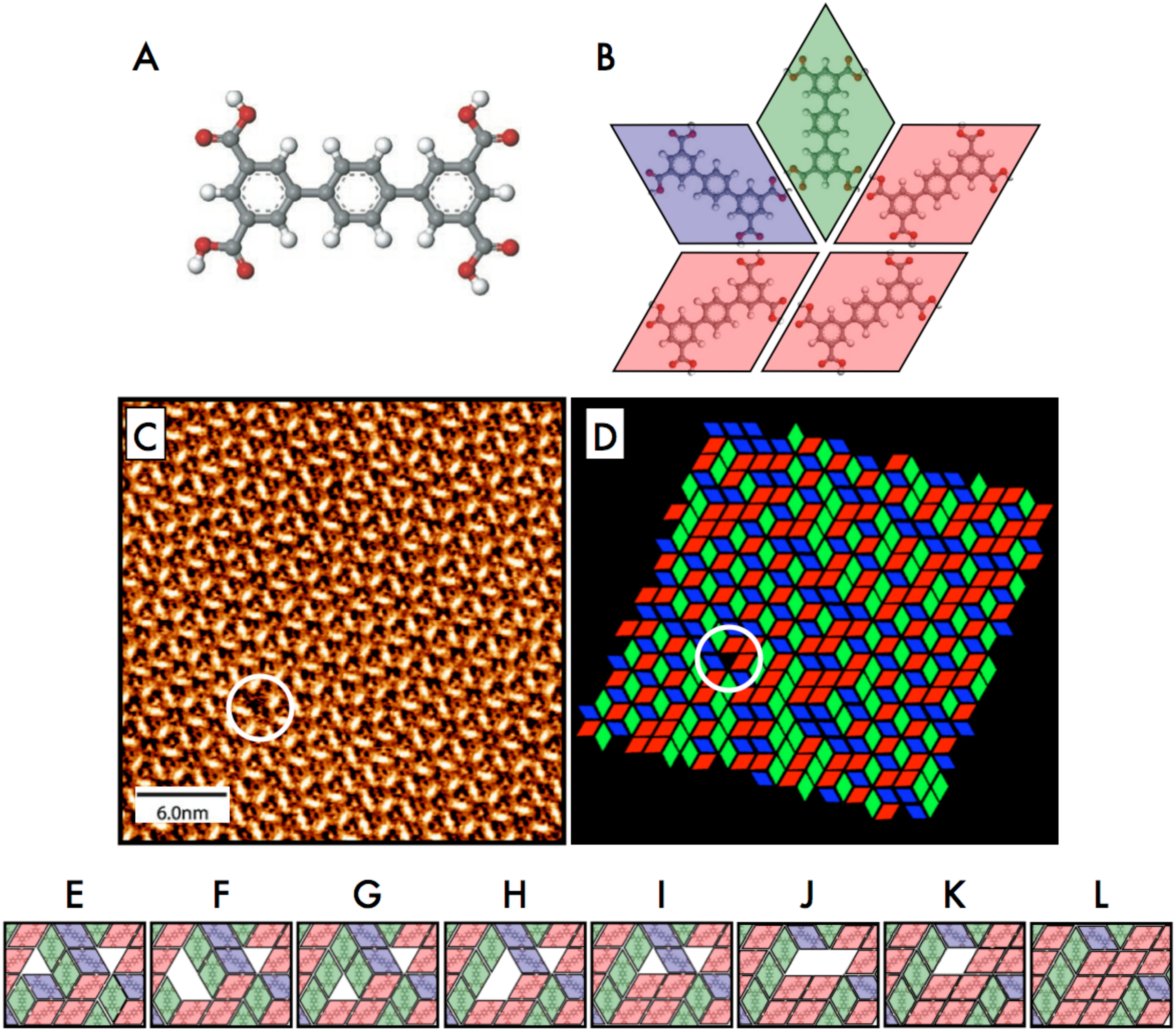}{({\bf A}) Molecular ball and stick diagram of p-terphenyl-3,5,3Õ,5Õ-tetracarboxylic acid (TPTC).  ({\bf B}) Example of an arrangement of TPTC molecules linked via hydrogen bonds when adsorbed on substrate, and rhombus tile representation; the tiles are coloured according to the three possible orientations of the molecule.  ({\bf C} and {\bf D}) Mapping to a Rhombus Tiling: (C) shows an STM image of a typical area of TPTC network adsorbed on graphite; the backbones of the TPTC molecules appear as bright rods in the image.  The corresponding rhombus tiling is shown in (D).  The molecular network is a rhombus tiling of the plane or, equivalently, a dimer covering of the honeycomb lattice.  The white circle shows the position of a tiling defect.  ({\bf E} to {\bf L}) Example of motion of tiling defects:  The leftmost defect (upward pointing triangle) effectively makes two hopping steps, between (E) and (G), and (G) and (I).  This motion is mediated by the desorption (F and H) and re-adsorption of a tile (G and I).  Between (I) and (K) the rightmost defect (downward pointing triangle) makes a step to the left, which brings it into contact with the leftmost defect.  The two annihilate with the adsorption of the last tile (L). (See Ref.\ \cite{Blunt} for details.)}{3ring}{hb}{.98}

\fig{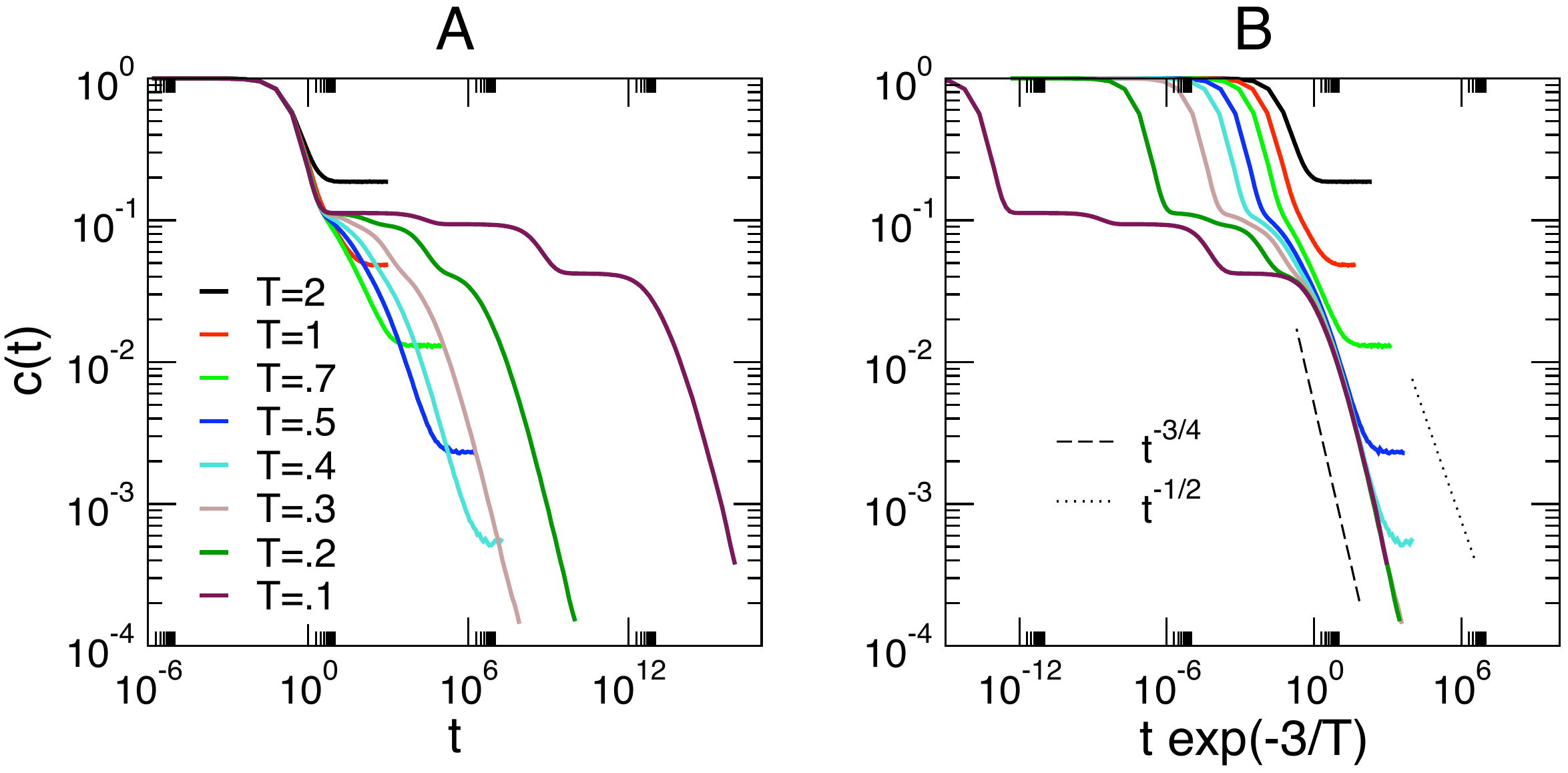}{({\bf A}) Relaxation of the concentration of defects $c(t)$ as a function of time, starting from an empty lattice, $c(0)=1$, for various temperatures $T$.  ({\bf B}) Same as (A), but time rescaled by the defect effective hopping rate $\Gamma \approx e^{-3/T}$.  The curves collapse at long times in this representation.  The dotted line indicates the power-law decay $(\Gamma~t)^{-1/2}$ expected from diffusion--pair-annihilation, $A+B \to 0$, in $d=2$.  The observed behaviour is closer to $c(t) \sim (\Gamma~t)^{-3/4}$, as indicated by the dashed line.}{ct}{h}{1}

\section{Model} 

We simulate a dimer covering of the honeycomb lattice, which is equivalent to a rhombus tiling of the plane  \cite{Fisher-Kasteleyn,Henley}.  That is, each rhombus tile is composed of an upward and a downward pointing triangle face-to-face; these triangles are centred at the sites of a honeycomb lattice, each in a different sublattice.  The dynamics we consider is one where the only possible moves are the adsorption of a tile on the lattice, if the two sites it would occupy are empty, or the desorption of a tile.  The number or tiles (and of tiling defects) is therefore not conserved.   This resembles the experimental situation where molecules are exchanged between substrate and solution \cite{Blunt}.  We only consider single-tile moves \cite{moves}.  We set the binding energy $J$ between neighbouring tiles to $J=1$.  For low temperatures, $T < 1$ (in units of $J$, and where $k_{\rm B}=1$), we simulate the dynamics using a version of Borz-Kalos-Lebowitz, or continuous-time, Monte Carlo \cite{BKL,Newman} which is particularly efficient for this problem.  We simulate systems of sizes $N=200 \times 200$ to $N=10^3 \times 10^3$ at all temperatures.

At low temperatures, when the density of tiles is high, desorption of tiles is rare.  The energy barrier to remove a tile surrounded by four neighbouring tiles is $\Delta E = 4$, and the rate for that transition is suppressed by a factor of $e^{-4/T}$.  A more likely transition is the removal of a tile neighbouring a defect, as the rate for this process scales as $e^{-3/T}$.  This will give rise to the effective propagation of tiling defects, as sketched in Figs.\ 1E-1L.  In Fig.\ 1E there are two tiling defects of opposite ``charge'' \cite{Henley}.  Fig.\ 1F shows the desorption of a tile next to the leftmost defect (a process of rate $\propto e^{-3/T}$) and Fig.\ 1G the subsequent adsorption of another tile (a process of rate $O(1)$, as it is energetically favourable).  The net effect is the hopping of the defect by one (sub)lattice site.  The sequence 1G-1I shows a second such step.  The effective hopping rate is therefore $\Gamma \propto e^{-3/T}$.

Defects of opposite charge can effectively react with each other.   This is sketched in Figs.\ 1I-IL.  In this case the rightmost defect hops one step to the left.  On meeting the opposite defect a gap large enough for a tile is formed, and the two defects are ``annihilated'' by the adsorption of a tile, at rate $O(1)$ as this is energetically favourable.  Of course, this process is reversible, and two opposing defects can be ``created'' by desorption of a molecular tile, at a rate $\propto e^{-4/T}$.   The effective dynamics of defects, therefore, resembles a reversible $A+B \leftrightarrow 0$ reaction-diffusion process \cite{Odor}, although, as we will see below, it is not clear that defect propagation is actually diffusive or that defect interaction can be approximated as being local (see also \cite{Bouttier,Jeng}).

\section{Results}

Figure 2A shows the evolution in time of the concentration of defects, $c(t)$, starting from an empty lattice at time zero, $c(0)=1$, at various temperatures $T$.   After a short initial transient of fast, temperature independent, tile adsorption, the system enters a regime of activated dynamics: most defects are isolated, and energy barriers need to be crossed for the dynamics to progress.  The dynamics becomes increasingly sluggish with decreasing temperature, and once times are long enough for defect motion to take place relaxation enters a scaling regime.  Fig.\ 2B shows that the rate limiting step is defect hopping: the long time data collapses if time is rescaled by the defect hopping rate, $t \to \Gamma t$.  The defect concentration decays as $c(t) \sim (\Gamma t)^{-\alpha}$, with $\alpha \approx 3/4$.  This exponent is somewhat different from the exponent $\alpha = 1/2$ of two-species diffusion-annihilation, $A+B \to 0$, in dimension two \cite{Odor,Toussaint}.  This could be an indication that defect propagation is non-diffusive, although an exponent of $\alpha = 3/4$ can also be explained by initial state fluctuations in the tiling case which differ from those of the standard $A+B \to 0$ problem \cite{alan}.   Eventually, for times $t \gg e^{4/T}$, the reverse process $0 \to A+B$ becomes accessible, and the concentration relaxes to its equilibrium value $c(t) \to c_{\rm eq}$.

\fig{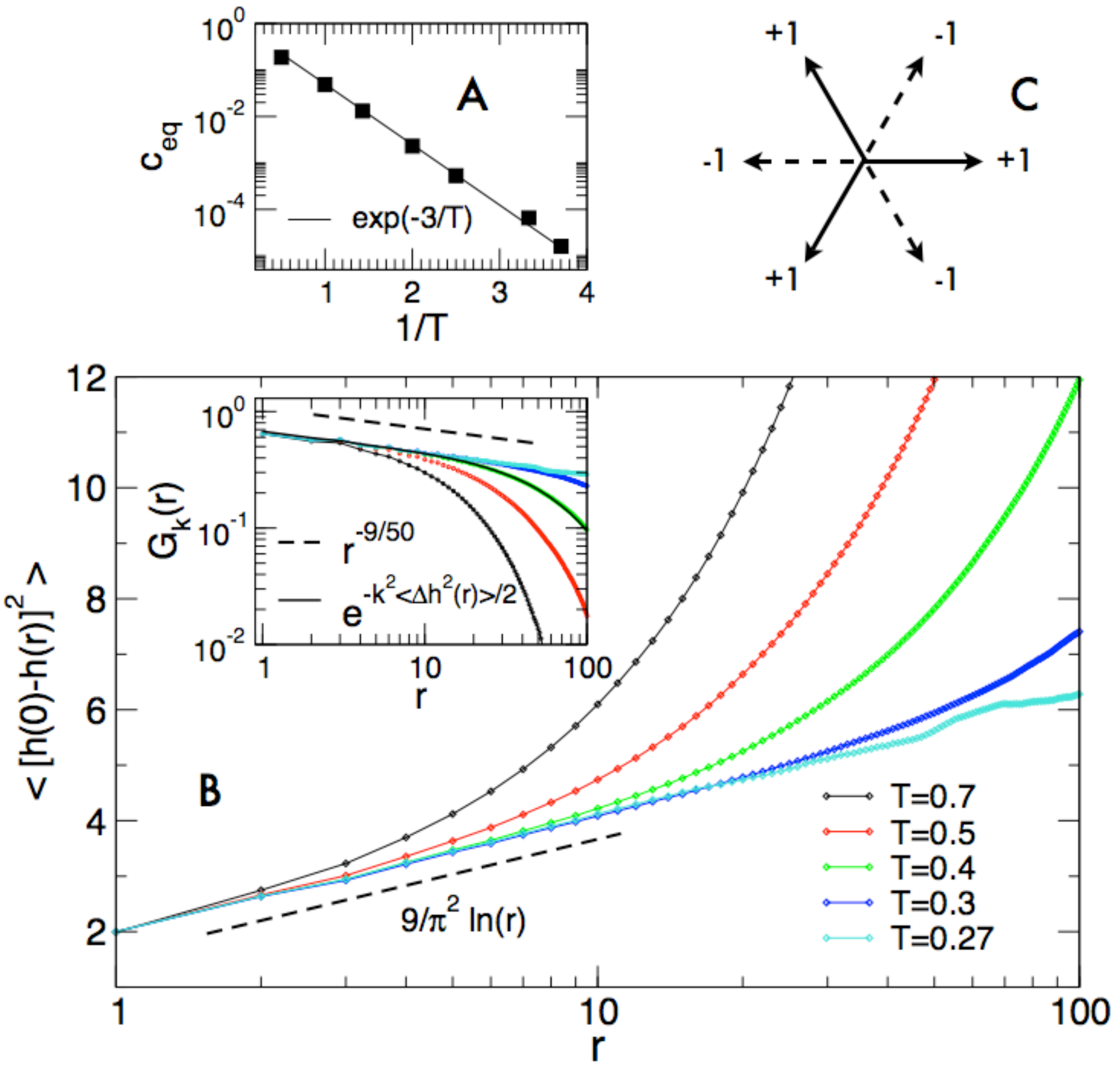}{({\bf A}) Equilibrium concentration of defects, $c_{\rm eq}$, as a function of temperature $T$.  The straight line corresponds to the fit $c_{\rm eq} = e^{-3/T}$.  ({\bf B}) Equilibrium height correlations at various temperatures.  The main panel shows $\langle [h(r)-h(0)]^2 \rangle$ as a function of distance $r$. As the defect concentration decreases with decreasing temperature, the curves approach the ideal tiling behaviour $\langle [h(r)-h(0)]^2 \rangle = 9/\pi^2 \ln{(r)}$.  The Inset shows the correlation function $\langle e^{i k \Delta h(r)} \rangle$, where $\Delta h(r)\equiv h(r)-h(0)$, for $k=\pi/5$; the correlation behaves in a similar way for other choices of $k$.  The dashed line is the power law behaviour for the ideal tiling, $r^{-9 k^2 / 2 \pi^2}=r^{-9/50}$ for this choice of $k$.  Once again, the lower the temperature, the longer the algebraic regime.  For one of the temperatures, $T=0.4$, we show that  $\langle e^{i k \Delta h(r)} \rangle = e^{-k^2/2 \langle [h(r)-h(0)]^2 \rangle}$, as expected from a Gaussian form of the free-energy for the height.  ({\bf C}) Scheme for obtaining the height representation of a tiling. A displacement along a tile edge leads to an increase in height by $+1$ or $-1$ as shown (cf. Fig.\ 1).  
}{eqcorr}{h}{1}

The equilibrium properties of the tilings are shown in Fig.\ 3.  The temperature dependence of the equilibrium defect concentration is given by $c_{\rm eq} \approx e^{-3/T}$, see Fig.\ 3A.  This is what one would obtain for a non-interacting gas of defects on the lattice with an energy cost of $E=3$ per defect.  
Fig.\ 3B shows the spatial correlations.  A rhombus tiling can be mapped to a height field on the triangular lattice \cite{Blothe}: the height $h$ changes by $\pm 1$ unit when traversing the edges between tiles, according to the prescription of Fig.\ 3C.  The main panel of Fig.\ 3B shows the height-height correlation function, $\langle [h(r)-h(0)]^2 \rangle$, as a function of distance $r$ (along lattice directions), for various temperatures \cite{height}.  At low temperatures this correlation approaches the ideal tiling limit $\langle [h(r)-h(0)]^2 \rangle = 9/\pi^2 \ln{(r)}$, corresponding to a Gaussian free-energy $F = \int d^2\vec{x} (K/2) |\nabla h(\vec{x})|^2$ for a continuous height field \cite{Henley}, with elastic constant $K=\pi/9$ \cite{Blothe}.   For finite $T$ the logarithmic behaviour is over a finite distance due to the presence of tiling defects \cite{Krauth}.  An alternative correlation function, $\langle e^{i k [h(r)-h(0)]} \rangle$ \cite{Kondev}, is shown in the Inset to Fig.\ 3B, for the specific value $k=\pi/5$ of the ``height space'' reciprocal vector \cite{Kondev} (the behaviour is similar for other choices of $k$).  At low temperatures the function becomes algebraic indicating long range correlations.  The decay exponent is close to $9 k^2 / 2 \pi^2$, as expected from the Gaussian form of the free-energy \cite{Kondev}.

\fig{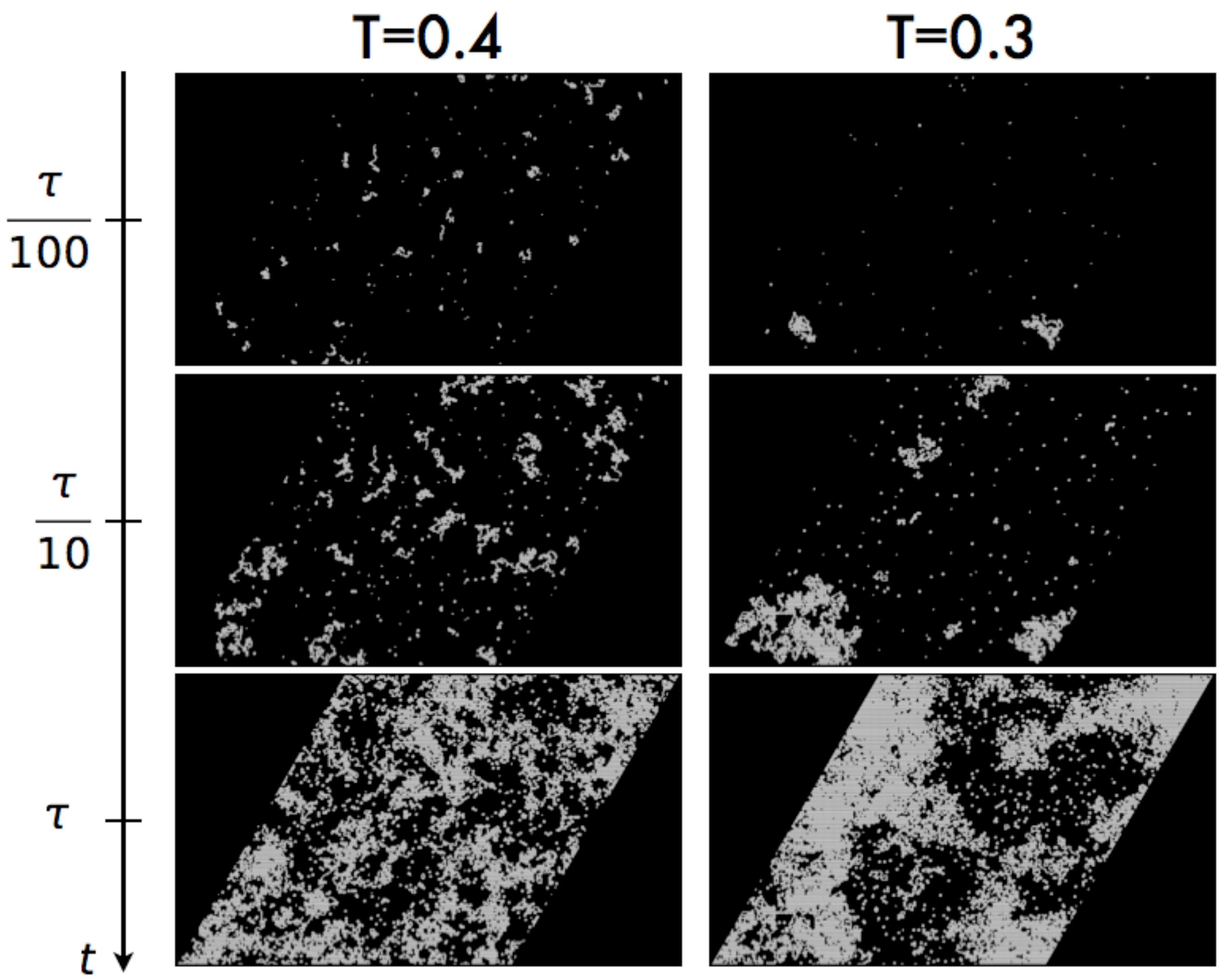}{Dynamic heterogeneity in random rhombus tilings.  The panels show the persistence field $P_i(t)$ of the local autocorrelation function $C_i(t)$ (see text) at various times $t$, for representative equilibrium trajectories, at two different temperatures $T$.  Black indicates $P_i(t)=1$, and white $P_i(t)=0$.  The average relaxation time is $\tau$, see Fig.\ 5.  Relaxation is clearly heterogeneous.  The size of dynamic heterogeneity grows with decreasing temperature.}{DH}{h}{1}

As described above, at low temperatures structural rearrangements are most likely in the neighbourhood of tiling defects.  This gives rise to heterogeneous relaxation, as illustrated in Fig.\ 4.  Here we plot the local autocorrelation function $C_i(t) \equiv \delta_{n_i(t),n_i(0)}$, where $n_i$ stands for the state of site $i$ in the lattice, say $n_i=0,1,2,3$ for empty, or occupied by a red, green or blue tile, respectively.  More precisely, Fig.\ 4 shows the corresponding persistence field, $P_i(t) = \prod_{t'=0}^t C_i(t)$: if site $i$ has never relaxed up to time $t$ then $P_i(t)=1$, and $P_i(t)=0$ otherwise.  The different panels show how relaxation is distributed in space at different times.  Clearly, the system relaxes in a heterogeneous, spatially correlated manner.  Fig.\ 4 also suggests that the size of these spatial dynamic correlations grows with decreasing temperature.  This is very similar to dynamic heterogeneity in structural glass formers \cite{reviewsDH}.

Figure 5 quantifies equilibrium relaxation and dynamic heterogeneity.  Fig.\ 5A shows the average (connected and normalised) autocorrelation function $C(t) \equiv \left( \langle C_i(t) \rangle - A \right) / \left( 1 - A \right)$, where $A \equiv \langle C_i(\infty) \rangle = c^2_{\rm eq} + (1-c_{\rm eq})^2/3$.  As expected, the autocorrelation function decays more slowly the lower the temperature.  The characteristic timescale for relaxation, $\tau$, obtained from these correlations is approximately $\tau = \tau_0 e^{\Delta/T}$, with $\Delta \approx 6.6$, see Inset to Fig.\ 5B.  Relaxation times thus increase with decreasing temperature following an Arrhenius law.  In the context of the glass transition this is often termed ``strong'' glass forming behaviour \cite{reviews}.  Moreover, the autocorrelations are close to exponential, rather than stretched exponential \cite{reviews}.  This could mean that relaxation is not collective, but Fig.\ 4 suggests otherwise.  The exponent $\Delta$ appears non-trivial: we have that $\Delta < 7$, the energy barrier to remove two neighbouring and parallel tiles (which would allow the adsorption of a distinct tile in the space created); this is the lowest energy barrier to purely local relaxation of the autocorrelation.  This indicates that relaxation is achieved more effectively through defect propagation, a collective mechanism.  Interestingly, we also have $\Delta > 6$, which is the value one would expect for diffusing defects, although this may simply be due to logarithmic corrections to diffusion in dimension two \cite{pers}.

\fig{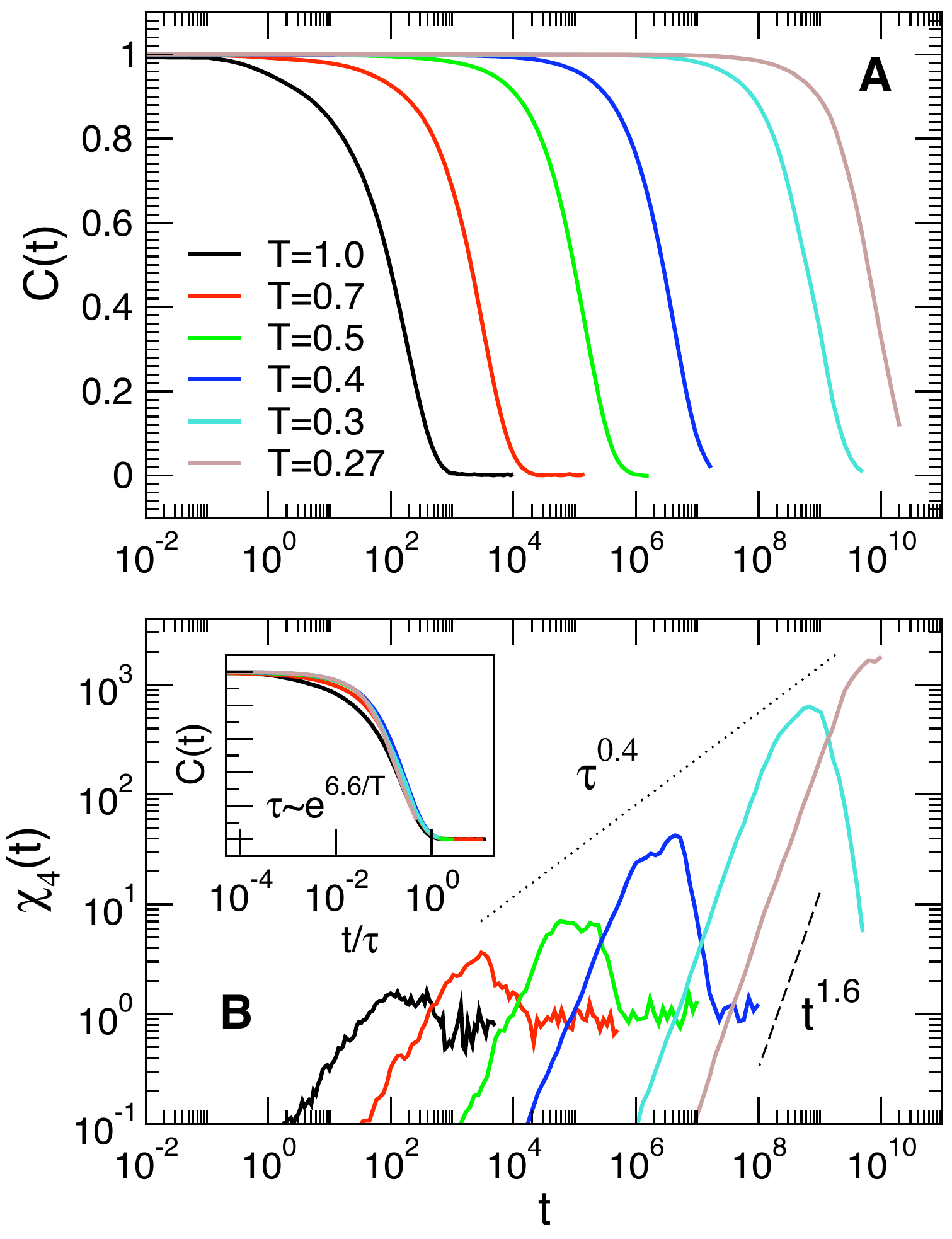}{({\bf A}) Average equilibrium autocorrelation function $C(t)$ at various temperatures $T$.  ({\bf B}) Four-point susceptibility $\chi_4(t)$ at the same temperatures as (A).  The four-point susceptibility has a maximum at $t \approx \tau$.  The peak value $\chi_4^{({\rm peak})} \sim t^\lambda$, with $\lambda \approx 0.4$, see dotted line.  The growth towards the peak follows the power law $\chi_4 \sim t^\mu$, with $\mu \approx 1.6$.  The Inset to (B) shows that the relaxation timescale for the autocorrelation function follows an Arrhenius law, $\tau \propto e^{\Delta / T}$, with $\Delta \approx 6.6$.
}{chi4}{h}{.8}

In Fig.\ 5B we show the ``four-point" susceptibility, $\chi_4(t) = N (1-A)^{-2} \left[ \langle \frac{1}{N^2} \sum_{ij} C_i(t) C_j(t) \rangle - C^2(t) \right]$, which measures sample to sample fluctuations in the correlator $C(t)$. This is an observable often used to quantify dynamic heterogeneity \cite{reviewsDH,Toninelli,chi4}.  $\chi_4$ is non-monotonic in time, peaking at times close to the structural relaxation time, $t \approx \tau$, where dynamic heterogeneity is most prominent.  The size of the peak of $\chi_4$ increases with decreasing temperature, indicating that dynamical fluctuations are larger at lower $T$.  The four-point susceptibility displays dynamic scaling, but again this is slightly different from what one expects from diffusing excitations in two dimensions \cite{Toninelli}: the peak value, $\chi^{\rm peak}_4=\chi_4(\tau)$, scales as a power of the relaxation time, $\chi^{\rm peak}_4 \sim \tau^\lambda$, with $\lambda \approx 0.4$ (rather than $\lambda=1$); and the growth of $\chi_4$ towards the peak goes as $\chi_4(t) \sim t^\mu$, with $\mu \approx 1.6$ (rather than $\mu=2$).

\section{Discussion}

We have shown that random tilings, of the kind corresponding to the experimental system of \cite{Blunt},  display features commonly associated with glass forming systems, most notably a pronounced slowdown at low temperatures and accompanying heterogenous relaxational dynamics.  In these nearly dynamically arrested tilings, structural relaxation occurs through the propagation of rare localised tiling defects.  This is an example of the mechanism of dynamic facilitation \cite{Fredrickson-Andersen,Garrahan}, whereby defects ``facilitate" molecular rearrangements in their immediate vicinity.  The fact that defects are scarce at low $T$, and that their motion is activated, leads to the observed slowdown, and to fluctuation-dominated, heterogeneous dynamics \cite{Garrahan}.  Relaxation by means of defect propagation-reaction is the hallmark of kinetically constrained models (KCMs) of glasses \cite{Ritort-Sollich}.  The dynamics of the random tiling systems studied here is very close to that of these idealised models.  For the simple rhombus tiling, when the number of tiles is not conserved, we have found Arrhenius timescales and exponential relaxation.  This is similar to the simpler KCMs, such as the Fredrickson-Andersen model \cite{Ritort-Sollich} in dimensions two or more \cite{Jack}.  The observed dynamic scaling properties may suggest however that defect dynamics is not simply that of diffusion-annihilation-creation, but this requires further exploration.

It is possible that more complex random tiling systems, such as those giving rise to quasicrystals \cite{Henley}, display even richer slow dynamics, in particular, super-Arrhenius timescales at low temperatures and stretched relaxation functions \cite{quasicrystals}.   Random tilings also offer a further testing ground for theories of the glass transition.  Specifically, it would be of interest to see whether their slow dynamics can be explained using the ``mosaic" perspective on glasses \cite{Xia,Bouchaud}.  Here we have a system with a finite configurational entropy density given by all the possible local tiling arrangements, the central tenet of the mosaic approach \cite{Xia,Bouchaud}.  The mismatch between mosaics however does not produce extended interfaces, but rather localised defects, which as shown above make the dynamics close to that of KCMs.  Can one therefore explain the structural relaxation of these systems via the entropic droplet picture of Refs.\ \cite{Xia,Bouchaud}? And could this highlight connections between the mosaic approach and that based on dynamic facilitation and KCMs \cite{cage}?  We hope to address these and other questions in future work.

\begin{acknowledgments}
This work was supported in part by EPSRC grant no.\ EP/D048761/1.
\end{acknowledgments}

\end{article}


\begin{thebibliography}{99}
\bibitem{bias} Due to a slight mismatch in the distance between the two end phenyl rings in a TPTC molecule, and that between phenyl rings across a carboxylicÐ-carboxylic hydrogen bond, the tilings of Ref.\ \cite{Blunt} are not ``ideal" \cite{Henley} as there is a small energetic penalty for the parallel tile arrangement.  They are ``interacting" random tilings \cite{Alet,Jacobsen}.  This difference in energy is however below the critical value at which such tilings would undergo a Kosterlitz-Thouless (KT) transition to an ordered phase \cite{Jacobsen}.
\bibitem{moves} In particular, we ignore moves where a cluster of tiles can be replaced without disturbing neighbouring tiles.  For example, in Figs.\ 1E-1G there is a central cluster of three tiles, red-blue-green, forming a hexagon of side one.  This triplet can be replaced by a green-red-blue combination without disturbing the surroundings.  This kind of moves are usually the ones considered for perfect tilings, see e.g. \cite{Henley2}.  In our case, they would be suppressed due to the energy cost of removing three tiles.  Furthermore, we do not consider moves where tiles ``slide" on the substrate, as this kind of process does not seem to apply to the experimental situation of Ref.\ \cite{Blunt} either.
\bibitem{height} 
An isolated defect in the tiling is a dislocation in the height representation, with Burgers vector of magnitude $+3$ ($-3$) for an upward (downward) pointing defect \cite{Kondev}.  When computing the height difference between two points in the tiling, along a lattice direction, if a defect is encountered we randomly choose to bypass it by the left or by the right.  This means that each defect adds or subtracts one unit of height randomly with equal probability.
\bibitem{pers} When relaxation is due to propagation of free defects, the persistence function is approximately given by $P(t) \approx e^{-c {\cal N}(t)}$, where $c$ is the concentration of defects, and ${\cal N}(t)$ is the average number of distinct sites visited by a defect in time $t$.  For diffusing defects in two dimensions, ${\cal N}(t) \sim (\Gamma t) / \ln{(\Gamma t)}$.  This, combined with the equilibrium concentration, $c_{\rm eq}$, gives, up to logarithmic corrections, $P(t) \sim e^{-t/\tau}$, with $\tau \sim c^{-1}_{\rm eq} \Gamma^{-1} \propto \exp{(6/T)}$.
\bibitem{quasicrystals} The relation between quasicrystals and certain glasses has been discussed recently in Ref.\  \cite{Keys}.  The slow dynamics of other kind of tilings is studied also in \cite{Shokef}.  Other random molecular networks with glassy characteristics have been reported in \cite{Otero}.
\end{thebibliography}
\end{document}